\documentclass[twocolumn,showpacs,preprintnumbers,amsmath,amssymb,prb]{revtex4}
\usepackage{graphicx, color}
\usepackage{dcolumn}
\usepackage{bm}

\begin{document}

\title{Enhancement of electronic inhomogeneity due to the out-of-plane disorder in Bi$_2$Sr$_2$CuO$_{6+\delta}$ superconductors observed by scanning tunneling microscopy/spectroscopy}

\author{A. Sugimoto$^1${\footnote{Present address: Graduate School of Integrated Arts and Sciences, Hiroshima University, Higashi-Hiroshima 739-8521, Japan.\\
Electronic address: asugimoto@hiroshima-u.ac.jp }}, S. Kashiwaya$^1$, H. Eisaki$^1$, H. Kashiwaya$^1$}
\author{H. Tsuchiura$^2$}
\author{Y. Tanaka$^3$}
\author{K. Fujita$^4$}
\author{S. Uchida$^{4,5}$}
\affiliation{$^1$Nanoelectronics Research Institute, National Institute of Advanced Industrial Science and Technology (AIST), 1-1-1 Umezono, Tsukuba 305-8568, Japan}
\affiliation{$^2$Department of Applied Physics, Tohoku University, Sendai 980-8579, Japan}
\affiliation{$^3$Department of Applied Physics, Nagoya University, Nagoya 464-8603, Japan}
\affiliation{$^4$Department of Advanced Materials Science, University of Tokyo, Tokyo 113-8656, Japan}
\affiliation{$^5$Department of Physics, University of Tokyo, Tokyo 113-0033, Japan}

\date{\today}
\begin{abstract}

Nanoscale electronic inhomogeneity in optimally doped single-layer Bi$_2$Sr$_{1.6}$L$_{0.4}$CuO$_{6+\delta}$, ($L$-Bi2201, $L$=La and Gd) as well as bilayer Bi$_2$Sr$_2$CaCu$_2$O$_{8+\delta}$ (Bi2212), superconductors has been investigated by the low-temperature scanning-tunneling-microscopy and spectroscopy. The inhomogeneous gap structures are observed above and below $T_c$ on Gd-Bi2201. Increasing the out-of-plane disorder which is controlled by mismatch of the ionic radius between Sr and $L$ results in the enhancement of pseudogap region with suppressed coherence peaks in the conductance spectrum, making the gap average larger and the gap distribution wider. The degree of electronic inhomogeneity measured by the distribution width of the gap amplitude is shown to have a correlation with $T_c$ in optimally doped Bi2201 and Bi2212.

\end{abstract}


\maketitle

To explore the mechanism of high-$T_c$ superconductors (HTSC's), tunneling spectroscopy is a powerful tool since it provides direct information on the electronic density of state.
Especially, recent experimental results of scanning tunneling microscopy and spectroscopy (STM and STS) have revealed the nanoscale electronic inhomogeneity which shows up as spatial modulation of gap amplitude. \cite{kmlang, matsuda_2p, howald_inhomo, sugiphysc04, kmcelroy}
The coherence peaks on both side of the gap edge are suppressed as the gap amplitude gets larger, which is similar to the pseudogap in normal state of underdoped cuprates. \cite{kmcelroy}
The inhomogeneous electronic structure is suggestive of quite unconventional superconducting state and gives a strong impact on the exploration of the mechanism of high-$T_c$ superconductivity.
Disorder introduced by the chemical doping processes is a possible origin of the electronic inhomogeneity. \cite{Ziqiang}
In fact, for Bi$_{2}$Sr$_{2}$CaCu$_{2}$O$_{8+\delta}$ (Bi2212), the doping is controlled by the amount of excess oxygen $\delta$ near the BiO layers whch is most likely a source of the microscopic electronic inhomogeneity. \cite{kmcelroy}

In addition to the oxygen nonstoichometry, the disorder introduced by cationic substitution is also expected to play an important role in HTSC.
It has been suggested that disorder outside the CuO$_2$ planes might be responsible for the large difference in $T_c$, ranging from 30 to 90 K, among different cuprates with a single CuO$_2$ plane in the unit cell.
Recently, Fujita $et \; al.$ have demonstrated that the cationic disorder in the layer next to the CuO$_2$ plane in single-layer cuprates substantially reduces $T_c$ without changing the carrier density. \cite{fujita}
However, at present it is not clear how cationic disorder outside the CuO$_2$ planes (out-of-plane disorder) is related to the microscopic electronic inhomogeneity.

In this paper, we report on a direct proof of the systematic variation of nanoscale electronic inhomogeneity observed by the STM and STS measurements on rare-earth-substituted Bi$_2$Sr$_2$CuO$_{6+\delta}$ ($L$-Bi2201) with the same doping level, which were well characterized in the preceding work \cite{fujita}.
The La or Gd ions are partially replaced for the Sr ion in the SrO plane, next to the CuO$_2$ plane.
The ionic radius of $L$ (1.22 \AA \;for La$^{3+}$ and 1.10 \AA \;for Gd$^{3+}$) is smaller than that of Sr$^{2+}$ ion (1.27 \AA), so that this mismatch introduces structural disorder in the SrO planes. Since the mismatch (and hence disorder) is larger in Gd-Bi2201, a comparison between La and Gd would sort out the correlation between out-of-plane disorder and electronic inhomogeneity in the CuO$_2$ plane. \cite{eisaki}
We show that the nanoscale gap inhomogeneity is strongly correlated with this out-of-plane disorder and the degree of inhomogeneity evaluated by gap amplitude distribution is intimately related to the superconducting transition temperature $T_c$.

The $L$-Bi2201 (nominally Bi$_2$Sr$_{1.6}$L$_{0.4}$CuO$_{6+\delta}$) single crystals were fabricated by the traveling-solvent-floating-zone (TSFZ) method. \cite{fujita}
$T_c$ is 34 K for La-Bi2201 and 14 K for Gd-Bi2201.
The samples were prepared to have optimal doping level by oxygen annealing under the same condition (at 650 $^{\circ}$C and for 48 h).
The constant carrier density is confirmed by the measurement of the transport properties, the slope of $T$-linear resistivity, and the magnitude of Hall coefficient. \cite{fujita}
For comparison, the optimally doped Bi2212 (nominally Bi$_{2.1}$Sr$_{1.9}$CaCu$_{2}$O$_{8+\delta}$, $T_c$=93 K) was also examined.
A homebuilt STM unit installed in a low-temperature chamber cooled by liquid He has been used for the present measurements. The temperature of the sample stage was typically fixed at 9 K except for temperature dependent measurements on Gd-Bi2201.
The single crystals were transferred to the STM unit soon after the cleavage inside the low-temperature chamber under the ultrahigh-vacuum atmosphere.
The Pt/Ir tip was cleaned by a high-voltage field emission process with Au film target just prior to the STM and STS observations.

Figure 1 shows STM-STS mesurements of temperature dependence on Gd-Bi2201 below and above $T_c$ (at 5 K, 9 K, and 26 K).
In Figs. 1(a)-1(c), we show (a) topography, (b) the spatial distribution of gap ($\Delta$ map) and (c) the spatial variation of the conductance spectra [$G(x, V) =dI/dV $] along the white line in Fig. 1(b) measured on Gd-Bi2201 at 5 K ($< T_c$). 
The gap $\Delta$ was defined as half of the maximum-conductance peak-to-peak voltage in each spectrum. 
In some spectra, the gap peak is suppressed in the occupied-state side (negative-bias side).
In these cases, $\Delta$ is defined by the peak voltage in the unoccupied-state side.
Figures 1(d)-1(f) show (d) topography, (e) $\Delta$ map, and (f) spatial variation of the conductance spectra at 9 K ($< T_c$).
At both  5 K and 9 K, the spatial inhomogeneous gap structures were clearly observed basically similar to these of Bi2212 as observed in previous works \cite{kmlang, kmcelroy, matsuda_2p, howald_inhomo} and our experimental results [see Fig. 2(d) and 2(g)].
 The comparison between the different samples will be discussed later.
At this stage, it is important to check the influence of the thermal broadening effect on the conductance spectra based on the temperature dependence measurement, since the temperature of material dependent measurements discussed later (9 K) is close to $T_c$ (=14 K) in the case of Gd-Bi2201.
However, as easily recognized by Figs. 1(c) and 1(f), the broadened gap peaks and the distribution of $\Delta$ (from 15 meV to 120 meV) at 9 K are quite similar to those of 5 K. Based on this fact, we conclude that the broadened spectra of Gd-Bi2201 were not due to the thermal effect but to some intrinsic characteristics of the electronic states such as the lifetime broadening due to the scattering by the disorder.
\begin{figure}[htbp]
\begin{center}
\includegraphics[width=8cm, height=13.81cm]{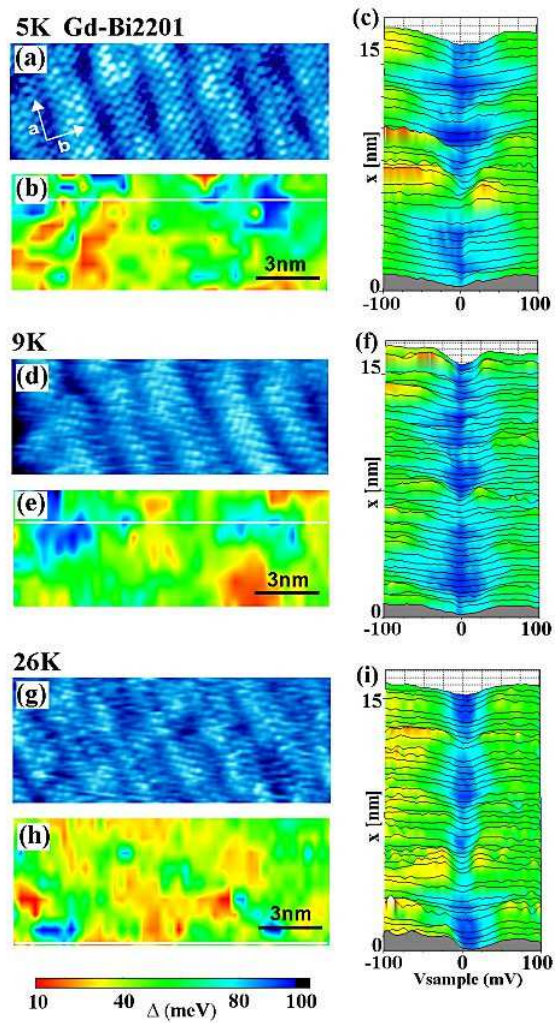}
\caption[]{{\footnotesize (Color online) The temperature dependence of STM and STS measurement on Gd-Bi2201.
(a) Topography at 5 K ($I_t=0.05$ nA, $V_{bias}=0.4$ V),
(b) $\Delta$ map at 5 K on the same area of (a) ($\bar {\Delta}$= 52 meV, $\sigma_\Delta > $ 22 meV within the field of view).
(c) The raw conductance spectra $G(x,V)$ along the white line in (b). 
(d) Topography at 9 K ($I_t=0.1$ nA, $V_{bias}=0.1$ V).
(e) $\Delta$ map at 9 K,
(f) The raw conductance spectra $G(x,V)$ along the white line in (e).
(g) Topography at 26 K ($I_t=0.1$ nA, $V_{bias}=0.1$ V).
(h) $\Delta$ map at 26 K.
(i) The raw conductance spectra $G(x,V)$ along the white line in (h).
The length-scale bars are common with topographies and z$\Delta$ maps.
The color scales of $\Delta $ are common with three $\Delta$ maps.
The magnitude of conductance is normalized by $G(x,0.1V)$. }}
\label{fig1}
\end{center}
\end{figure}
Moreover, we should note that the inhomogeneous gap structures survive even above $T_c$.
In Figs. 1(g)-1(i), we show (g) topography, (h) $\Delta$ map, and (i) spatial variation of the conductance spectra at 26 K ($>T_c$).
The spectral features above $T_c$ and the spatial extent of the inhomogeneity of the $\Delta$ are again similar to those observed at 5 K and 9 K.
These results indicate that the inhomogeneous electronic structure and the conductance spectra are relatively insensitive to the temperature up to 26 K even to the onset of superconductivity.
This implies that most of the gap structures observed here can be attributed to the pseudogap origin rather than the superconducting origin.
The existence of the gap structure above $T_c$ is consistent with the previous results of STS,\cite{kugler} angle-resoleved photoemission spectroscopy\cite{harris_arpes} (ARPES) and NMR (Ref. \cite{guo_nmr}) experiments on Bi2201.

From our experimental results of the $\Delta$ inhomogeneity above $T_c$, it is necessary to reconsider the meaning of the $\Delta$ inhomogeneity including that below $T_c$.
The degree of $\Delta$ inhomogeneity (for example, the standard deviation of $\Delta$) may not be directly attributed to superconductivity. 
However, the $\Delta$ inhomogeneity generated by pseudogap may play an important role as the background of superconductivity and has some correlation with $T_c$.
Anyway, at present, although the meaning of the degree of $\Delta$ inhomogeneity is ambiguous, it has some significant features for high $T_c$ superconductivity.
The relation between the deviation of $\Delta$ and $T_c$ measured by microscopic probe of STM and STS is considered to be corresponding to the relation between the mismatch of ion radius and $T_c$ measured by macroscopic method. \cite{fujita}
\begin{figure}[htbp]
\begin{center}
\includegraphics[width=8cm, height=15.2cm]{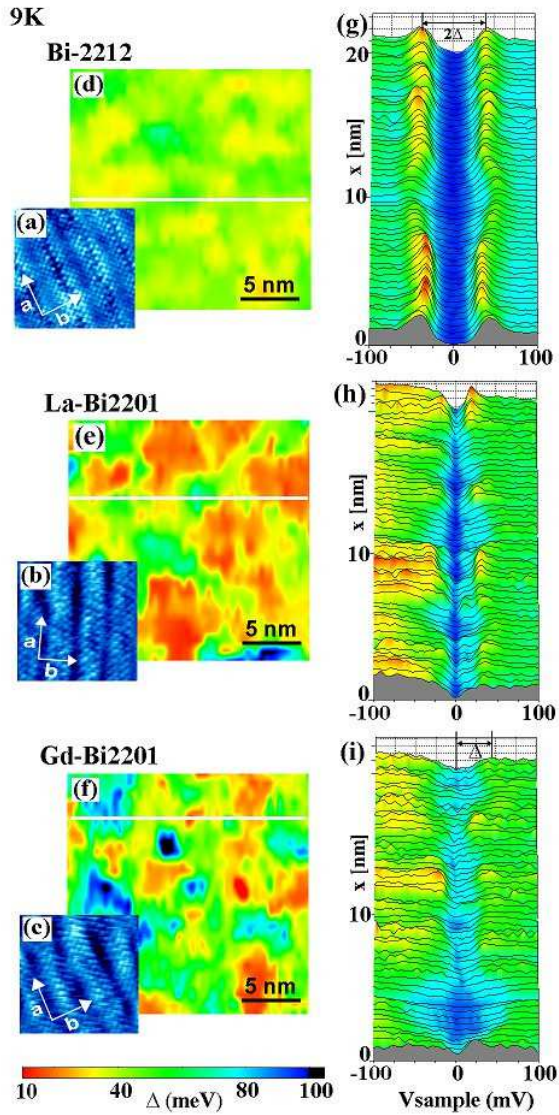}
\caption[]{{\footnotesize (Color online) (a)-(c) Topographic images on (a) Bi-2212 ($I_t=0.1$ nA, $V_{bias}=0.07$ V), (b) La-Bi2201 ($I_t=0.1$ nA, $V_{bias}=0.1$ V), and (c) Gd-Bi2201 ($I_t=0.1$ nA, $V_{bias}=0.1$ V) at 9 K. 
(d)-(f) $\Delta$ maps of 20 nm $\times$ 20 nm area on (d) Bi2212. (e) La-Bi2201. (f) Gd-Bi2201 at 9 K, including the topography area in (a), (b), and (c), respectively. The length-scale bars are common with topographies and $\Delta$ maps.
(g)-(i) The raw conductance spectra $G(x,V)$ along the white line in each $\Delta$ maps at 9 K. The magnitude of conductance is normalized by $G(x,0.1V)$. }}
\label{fig2}
\end{center}
\end{figure}

Figures 2(a)-2(c) show the topographic images measured on (a) Bi2212, (b) La-Bi2201, and (c) Gd-Bi2201 at 9 K, respectively.
In each topography, supermodulation in the BiO layer is observed.
The whole area in each sample is covered with these supermodulation structures with periodic length of 2.5$\pm$0.4 nm at least within an area of 70 nm $\times$ 70 nm, confirmed by the observation of wide-area topography.
The condition of the exposed BiO layer is considered to have no drastic difference between Bi2212 and $L$-Bi2201.
Figures 2 (d)-(f) are the $\Delta$ maps measured on (d) Bi2212, (e) La-Bi2201 and (f) Gd-Bi2201 within the area of 20 nm $\times$ 20 nm.
Figures 2(g)- 2(i) are spatial variations of the conductance spectra measured along the white line in each $\Delta$ map. 
  In the gap map of Bi2212, one can easily see that $\Delta$ varies on a nanometer scale as observed in previous works. 
The spatial distributions of $\Delta$ on La- and Gd-Bi2201 are basically similar to that of Bi2212 but the distribution of the gap amplitude is much broader.
It is seen that, as disorder becomes stronger, in the order of Bi2212, La-Bi2201, and Gd-Bi2201, the fraction of large-gap area (dark colored area) increases.
The length scale of the inhomogeneity (2$\xi_{\Delta}$) defined as the full width of half maximum of autocorrelation analysis is $2\xi_\Delta \simeq$2.7 nm for La-Bi2201, 2.2 nm for Gd-Bi2201, and 3.1 nm for Bi2212.
It is slightly smaller in Gd-Bi2201.
These lengths are comparable with the superconducting coherence length $\xi_0$ of the order of a few nm.
Note that, if $\xi_{\Delta}$ follows the BCS relationship $\xi_0=\hbar v_F/\pi\Delta_0 \propto 1/T_c $, $\xi_{\Delta}$ of Gd-Bi2201 should be longer by a factor of 2 or more than that of La-Bi2201.
Therefore, $\xi_{\Delta}$ does not scale with $T_c$, but it is roughly the same among Bi-based cuprates.

To see the evolution of the spectrum with increasing disorder the conductance spectra are shown in Figs. 3(a)-3(c).
Each curve is obtained by averaging all the spectra with identical gap amplitude within the $\pm$ 2 meV in each sample.
Compared with the spectra of Bi2212, the spectra of $L$-Bi2201 exhibit following distinct features.

(1) $\Delta$ is more widely distributed in the range between 10 and 100 meV, and the larger gap region ($\Delta>$50 meV) dominates. For large gaps, the spectrum has V-shaped structure without showing sharp coherence peaks. 

(2) The zero-bias conductance (ZBC) is appreciably large and increases as $\Delta$ decreases. The ZBC in Gd-Bi2201 is larger than that in La-Bi2201.

The distributions of $\Delta$ at 9 K are shown in histograms in Fig. 3(d), Bi2212, (e), La-Bi2201, and (f), Gd-Bi2201.
One can see that the distributions of $\Delta$ in La- and Gd-Bi2201 are much broader, and show a long tail in the high-energy side.
The parameters that characterize the $\Delta$ distribution ($\bar{\Delta}$ and $\sigma_{\Delta}$, the spatial averaged gap and the standard deviation, respectively) are $\bar{\Delta}$=42 meV and $\sigma_{\Delta}$=7 meV for Bi2212, $\bar{\Delta}$=36 meV and $\sigma_{\Delta}$=17 meV for La-Bi2201, and $\bar{\Delta}$=48 meV and $\sigma_{\Delta}$=22 meV for Gd-Bi2201.
If the inhomogeneity of $\Delta$ is simply generated by the spatial variation of local carrier density, the average gap amplitude $\bar{\Delta}$ should be identical between La- and Gd-substituted systems, since they have the same doping level. \cite{fujita}
\begin{figure}[htbp]
\begin{center}
\includegraphics[width=8cm, height=9.33cm]{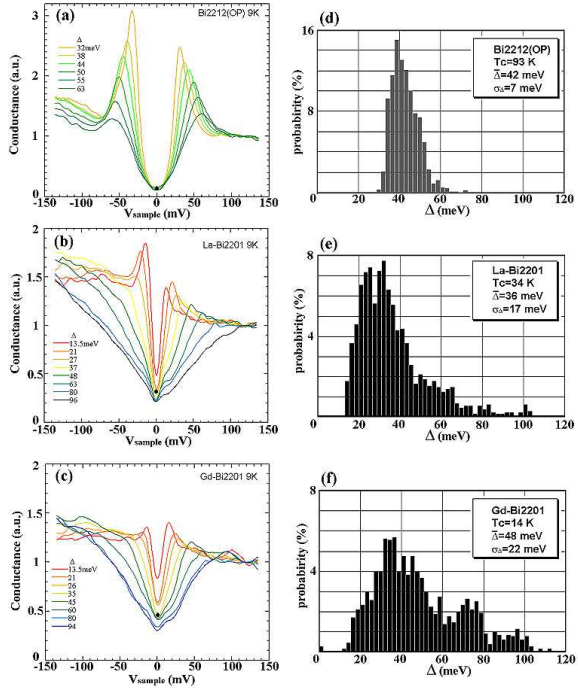}
\caption[]{{\footnotesize (Color online) (a)-(c) The $\Delta$-resolved spectra on (a) Bi2212, (b) La-Bi2201, and (c) Gd-Bi2201 at 9 K.
The spectra were obtained by collecting and averaging from all the raw spectra data within the $\pm$ 2 mV of each $\Delta$, except for the case of $\Delta \ge$ 80 meV within the $\pm$ 4 meV. The black dot in each zero-bias position indicates the average of zero-bias conductance. (d)-(f) The $\Delta$ histogram on (d) Bi2212, (e) La-Bi2201, and (f) Gd-Bi2201 at 9 K.}}
\label{fig3}
\end{center}
\end{figure}
However, in Gd-Bi2201, $\bar{\Delta}$ is larger than that in La-Bi2201, reflecting the fact that the fraction of large gap region increases in the former.
This indicates that disorder in the SrO layers influences the $\Delta$ distribution and the average value $\bar{\Delta}$ in Bi2201 and that $\bar{\Delta}$ is not directly related to $T_c$ nor to the total carrier density.
However, it is still an open question whether or not disorder gives rise to modulation of local carrier density.

Note that the conductance spectra with identical gap amplitude are significantly different between La-Bi2201 and Gd-Bi2201.
This is in contrast to the case of Bi2212 with different doping levels, \cite{kmcelroy} where the conductance spectrum at the position of identical $\Delta$ is found to be nearly the same for all doping levels.
This difference as well as the appreciable ZBC may arise from much stronger disorder in Bi2201.
The large-gap spectrum with V-shaped structure is similar to that observed for heavily underdoped Bi2212(Refs. \cite{kmlang, kmcelroy}) and lightly doped Na$_x$Ca$_{2-x}$CuO$_2$Cl$_2$ (Refs. \cite {kohsaka, hanaguri}), although the spectrum shape is slightly different from that for underdoped cuprates, possibly due to different location of the Van Hove singularity. The different spectral features including ZBC in Bi2201 will be discussed in more details elsewhere. \cite{qpl}

We see that $\bar{\Delta}$ of Bi2212 ($\bar{\Delta}$=42 meV) is higher than that of La-Bi2201($\bar{\Delta}$=36 meV) while lower than that of Gd-Bi2201 ($\bar{\Delta}$=48 meV).
There is no apparent correlation between $\bar{\Delta}$ and $T_c$. 
On the other hand, $T_c$ is found to be scale with $\sigma_{\Delta}$, as displayed in Fig. 4, in which $T_c$ is plotted as a function of $\sigma_{\Delta}$ including the available data for Bi2212 near optimal doping. \cite{kmlang, kmcelroy, hudson_ni, kinoda_apl}
For Bi2212, we plot a few data measured on nominally overdoped samples, but their $T_c$ values are so high ($T_c > 82$ K) that their doping level is considered to be near optimal.
$T_c$ is found to decrease linearly with of $\sigma_{\Delta}$ over a wide range of $T_c$ values, irrespective of the number of the number of CuO$_2$ planes and of the dominant disorder sites.
In Bi2212, disorder is presumably excess oxygen atoms near the BiO layers, but cationic disorder also exists in the SrO layers in the form of Bi$_{2+x}$Sr$_{2-x}$CaCu$_2$O$_{8+\delta}$.
The resent STM-STS observation by McElroy $et\;al.$ \cite{kmcelroy2} suggests that the excess oxygen atoms are a dominant source of gap inhomogeneity in Bi2212.
On the other hand, cationic disorder in SrO layers dominates in Bi2201 and is probably responsible for much broader gap distribution and maybe for lower $T_c$

The inset of Fig. 4 illustrates a schematic variation of $T_c$ as functions of doping and disorder (the present experimental conditions are indicated by open circles).
We have shown that the spatial distribution of $\Delta$ as well as the averaged $\Delta$ is largely different among the samples with different degrees of disorder but with constant doping level.
The systematic decrease of $T_c$ with increasing disorder at fixed doping level indicates that out-of-plane disorder, rather than local modulation of the carrier density, is a dominant factor that influences $T_c$.
This is compatible with the recent STM-STS result by McElroy $et\;al.$ \cite{kmcelroy2} on Bi2212 with excess oxygen atoms as a source of disorder.
They conclude that disorder does not change the carrier density in the CuO$_2$ plane locally.
\begin{figure}[htbp]
\begin{center}
\includegraphics[width=7cm, height=6.24cm]{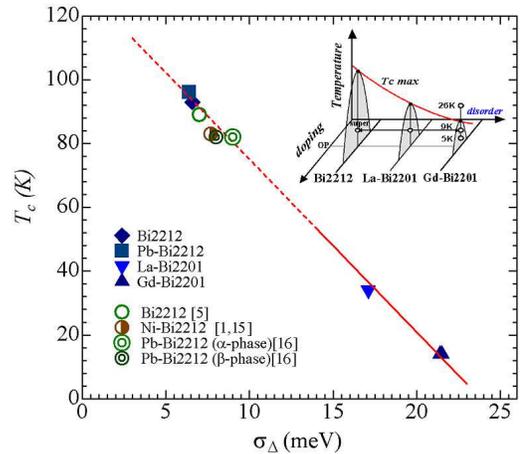}
\caption[]{{\footnotesize (Color online) The variation of $T_c$ as the function of $\sigma_{\Delta}$. The fitting line is obtained from the data of optimal doped Bi2212 and $L$-Bi2201 in our experiments. The plots of underdoped and overdoped Bi(Pb)2212 include the data from other reports (Refs. \cite{kmlang, kmcelroy, hudson_ni, kinoda_apl}). The inset shows the phase diagram of Bi-based HTSC with the axis of ``disorder."}}
\label{fig4}
\end{center}
\end{figure}
A recent calculation shows that out-of-plane disorder, such as dopant oxygen atoms, could locally modulate the pairing interaction and work as impurities for off-diagonal scattering which gives rise to nanoscale modulation of $\Delta$ as well as to the suppression of the coherence peaks. \cite{nunner}
However, it is not clear whether strong disorder such that in Bi2201 may give arise to appreciable ZBC.

In summary, a very broad gap distribution is observed by the STM and STS in the superconducting state of the disorder-controlled single-layer cuprate Bi2201.
Inhomogeneous gap structures were observed above and below $T_c$.
As the out-of-plane disorder increases, the ``pseudogap" region with larger gap and suppressed coherence peaks increases while the superconducting region decreases.
For the Bi-based cuprates including bilayer Bi2212, the degree of electronic inhomogeneity measured by the width of gap distribution $\sigma _\Delta$ is found to show a linear correlation with $T_c$ when the doping level is kept near optimal.
The result is suggestive of a possibility that the reduction of $T_c$ with increasing $\sigma_\Delta$ is due to the reduction of $\rho_s$. [In fact, it has been confirmed by the preliminary the muon relaxation rate ($\mu$SR) measurement that $\rho_s$ in Gd-Bi2201 is lower than that in La-Bi2201(Ref. \cite{fujita})].
In this regard, the decrease of $T_c$ with increasing $\sigma _{\Delta}$ is reminiscent of the relation between $T_c$ and $\rho _s$ determined by $\mu$SR, so called ``Uemura-plot."\cite{uemura}
Thus, we speculate that the low $T_c$ in Bi2201 is owing to strong disorder and $T_c$ may be considerably improved if the disorder is reduced.

This work was supported by Grant-in-Aid for Japan Society for Promotion of Science (JSPS) and by a Grant-in-Aid for Scientific Research in Priority Area from MEXT, Japan.\\

$^*$ Present address: Graduate School of Integrated Arts and Sciences, Hiroshima University, Higashi-Hiroshima 739-8521, Japan.\\
Elecronic address: asugimoto@hiroshima-u.ac.jp

\end{document}